\documentclass[letterpaper]{article} 
\usepackage{aaai25}  
\usepackage{times}  
\usepackage{helvet}  
\usepackage{courier}  
\usepackage[hyphens]{url}  
\usepackage{graphicx} 
\usepackage{natbib}  
\usepackage{caption} 
\usepackage{algorithm}
\usepackage{algorithmic}
\usepackage{amsmath}

\usepackage{xcolor}
\newcommand{\answerYes}[1]{\textcolor{blue}{#1}} 
 
\newcommand{\answerNA}[1]{\textcolor{gray}{#1}}

\usepackage{newfloat}
\usepackage{listings}
\DeclareCaptionStyle{ruled}{labelfont=normalfont,labelsep=colon,strut=off} 
\floatstyle{ruled}
\newfloat{listing}{tb}{lst}{}
\floatname{listing}{Listing}
\title{Cyber Food Swamps: Investigating the Impacts of Online-to-Offline\\ Food Delivery Platforms on Healthy Food Choices}

\author{
    Yunke Zhang\textsuperscript{\rm 1}, Yiran Fan\textsuperscript{\rm 2}, Peijie Liu\textsuperscript{\rm 1}, Fengli Xu\textsuperscript{\rm 1}\thanks{Corresponding Authors}, Yong Li\textsuperscript{\rm 1$\ast$}\\
    }

    \affiliations{\textsuperscript{\rm 1}Department of Electronic Engineering, Beijing National Research Center for Information Science and Technology (BNRist), Tsinghua University\\
    \textsuperscript{\rm 2}Department of Computer Science and Engineering, University of California San Diego,\\
        zyk21@mails.tsinghua.edu.cn, yif032@outlook.com, liupj23@mails.tsinghua.edu.cn, fenglixu@tsinghua.edu.cn, liyong07@tsinghua.edu.cn
    }

\begin{document}

\maketitle

\begin{abstract}
Online-to-offline (O2O) food delivery platforms have greatly expanded urban residents' access to a wide range of food options by allowing convenient ordering from distant food outlets. However, concerns persist regarding the nutritional quality of delivered food, particularly as the impact of O2O food delivery platforms on users’ healthy food remains unclear. This study leverages large-scale empirical data from a leading O2O delivery platform to comprehensively analyze online food choice behaviors and how they are influenced by the online exposure to fast food restaurants, \emph{i.e.,} online food environment. Our analyses reveal significant variations in food preferences across demographic groups and city sizes, where male, low-income, and younger users are more likely to order fast food via O2O platforms. Besides, we also perform a comparative analysis on the food exposure differences in offline and online environments, confirming that the extended service ranges of O2O platforms can create larger ``cyber food swamps''. Furthermore, regression analysis highlights that a higher ratio of fast food orders is associated with ``cyber food swamps'', areas characterized by a higher proportion of accessible fast food restaurants. A 10\% increase in this proportion raises the probability of ordering fast food by 22.0\%. Moreover, a quasi-natural experiment substantiates the long-term causal effect of online food environment changes on healthy food choices. These findings underscore the need for O2O food delivery platforms to address the health implications of online food choice exposure, offering critical insights for stakeholders aiming to improve dietary health among urban populations.
\end{abstract}

\begin{links}
\link{Code \& Dataset}{https://github.com/tsinghua-fib-lab/CyberFoodSwamp}
\end{links}

\section{Introduction}

Over the past decade, with the advancement of technologies of mobile commerce and crowd-sourcing platforms, we have witnessed a surge in the adoption of online-to-offline (O2O) food delivery~\cite{zhao2021o2o, shroff2022online, meemken2022research}. As of 2023, the size of O2O food delivery user in China reached 545 million, accounting for 50\% of the total netizen population of the nation~\cite{2023report}. Through the information matching and dissemination service provided by O2O food delivery platforms, users can access meals from any desired restaurants, thereby reshaping the dining habits of numerous urban residents.

Although the convenience of O2O food delivery is undeniable, concerns persist regarding the nutritional quality of delivered food, with numerous criticisms pointing to its generally low nutritional value~\cite{dai2022nutritional}. Drawing on the concept of ``food swamps'', which describes areas oversaturated with unhealthy food outlets in the physical world~\cite{stowers2017food}, O2O food delivery platforms have the potential to expand food access and create virtual food swamps that shape consumer choices towards unhealthy restaurants, leading to adverse health outcomes~\cite{stephens2020food}. Therefore, it is essential to assess how online food environments on these platforms influence users' healthy food choices.

Existing studies primarily used surveys and field studies to evaluate O2O food delivery platforms from the perspective of healthiness, including the nutritional value of delivered food~\cite{brar2021geographic, dai2022nutritional}, consumer perceptions of healthy food availability~\cite{dai2022nutritional, eu2021consumers, osaili2023healthy}, and healthy food choices~\cite{osaili2023healthy, giacomini2024variables}. These studies often suffered from limitations in the scale of research samples and lack a direct link between online food environment and consumers' real-world food choices. Furthermore, while significant efforts have been made to examine the impact of traditional dining environments on food choices~\cite{feng2010built, althoff2022large, garcia2024effect}, their findings may not directly correspond to the effects of online food environments. This is particularly relevant as O2O food delivery significantly expands the spatial range of food options, highlighting the need to compare offline and online food swamps.

In this study, we aim to provide comprehensive evidence of the healthiness of online food environments and its impact on healthy food choices using large-scale empirical O2O food delivery consumption data through following three research questions:

\textbf{RQ1}: How healthy are online food environments? Is there overall healthiness discrepancy of O2O food delivery consumption?

\textbf{RQ2}: What are the differences between offline and online food environments?

\textbf{RQ3}: How does online food environment impact consumers' food choices? Are these effects consistent across demographic groups?

To answer the above questions, we conduct comprehensive data-driven analyses and derive insightful findings based on O2O food delivery restaurant and consumption data at both city and individual levels, collected from one of the largest O2O food delivery platforms in China. First, we focus on the macro-level health discrepancies within the O2O food delivery industry. We uncover a scaling law that governs the relation between O2O food delivery consumptions and city sizes, showing larger cities have significantly higher proportions of fast food restaurants and placed delivery orders, which may be attributed to the faster pace of social life in these cities~\cite{bettencourt2007growth}. Healthiness of food choices is also correlated with demographic, with males, individuals with lower incomes, and younger people exhibiting a significantly higher likelihood of ordering fast food online. This trend, derived from large-scale empirical data, mirrors previous findings on food destinations based on small-sample surveys~\cite{jacqueline2012predictors}. Second, we compare the distributional patterns between offline and online food environments. The number of accessible O2O restaurants is significantly higher than that of local offline restaurants, while the healthiness of the online food environment is slightly better than that of the offline environment. However, the extended service range of O2O platforms expands the spatial coverage of food swamps. The online food environment shows a stronger correlation with healthy food choices on O2O food delivery platforms. Finally, following the analytic framework proposed by Garc{\'\i}a Bulle Bueno \textit{et al.}~\cite{garcia2024effect}, we examine the relationship between online food environments and healthy food choices. Logistic regressions reveal that a higher proportion of fast food restaurants in a user’s location is associated with a higher likelihood of ordering fast food. Notably, low-income individuals and younger users are more affected by this environment, underscoring the heterogeneous susceptibility of users on online platforms~\cite{li2022exploratory, sukiennik2024uncovering}. A quasi-natural experiment based on users who permanently change their ordering context further substantiates the gradual, negative causal impact of online food environments dominated by unhealthy fast food options on healthy dining habits, highlighting the urgent need for regulations addressing ``cyber food swamps''.

Our main contributions can be summarized as follows. \\
$\bullet$ We leverage empirical O2O food delivery consumption data to comprehensively understand the healthiness of the online food environment and its impact on healthy food choices, highlighting the significant potential for exploring the interplay between the web and society using data-driven methods. \\
$\bullet$ We demonstrate macro-level health discrepancy within the online food environment by uncovering city-level scaling laws of O2O food delivery adoption and identifying disparities in healthy food choices across demographic groups. \\
$\bullet$ We compare the healthiness of offline and online food environments and confirm the amplifying effect of O2O food delivery on real-world food swamps. \\
$\bullet$ We reveal the effects of online food environments on healthy food choice behaviors through a quasi-intervention experiment utilizing a detailed, individual order-level dataset, indicating the negative health impacts of ``cyber food swamps''.

\section{Related Works}
We categorize the related works into three dimensions.

\subsubsection{O2O Food Delivery Adoption and Health Perception}

Many previous studies and reports have substantiated the rapid growth of O2O food delivery in global cities, highlighting it as a convenient and viable solution to address limited access to healthy food options~\cite{dillahunt2019online, yeo2017consumer}. During the COVID-19 pandemic, social distancing measures made O2O food delivery a primary means of obtaining food, further accelerating the industry's growth~\cite{dsouza2021online, li2020review, hong2021factors}. At the same time, the nutritional and health implications of O2O food delivery have attracted increasing attention from researchers worldwide. As for users' perceptions of delivered food, surveys conducted in Jordan~\cite{osaili2023healthy}, Malaysia~\cite{eu2021consumers}, and China~\cite{dai2022nutritional} found that most participants believe food delivered through O2O food delivery is generally less healthy than food served in restaurants. However, many of these participants still report a high reliance on such services. Several studies have evaluated the healthiness of O2O food delivery meals through sampling. An analysis of food delivery menus in Canada revealed that the majority of menus scored below the standards set by the Healthy Eating Index~\cite{brar2021geographic}. Similarly, a study on the nutritional quality of food sold on Chinese O2O food delivery platforms found that most popular items have low nutritional value~\cite{dai2022nutritional} -- younger consumers, in particular, tend to pay less attention to the nutritional content of delivered meals, which can lead to health issues such as elevated cholesterol and obesity with long-term consumption. In Italy, an analysis of differences in willingness to use food delivery services among users found that individuals with lower health literacy are more dependent on O2O food delivery~\cite{giacomini2024variables}. Most current studies are limited to small-scale data collection through surveys or sampling, without assessing the overall health preferences of O2O food delivery users across a broader population. In our study, we leverage a large-scale empirical dataset covering the consumption behaviors of hundreds of thousands of users to comprehensively analyze their online food environments and healthy food choices.

\subsubsection{Health Impacts of Food Environment}

Extensive efforts have been made to investigate the impact of healthy food accessibility on residents' food choices and health outcomes, which can be categorized into three stages based on data sources. Initially, researchers relied on interviews and field surveys to understand individuals' eating habits and the dining environment in their residential areas. However, the limited scale of these studies made it difficult to conduct comprehensive research across different regions, resulting in mixed and non-generalizable findings. For example, some studies reported an association between high exposure to fast food outlets and residents' unhealthy diets and obesity rates~\cite{wang2008changes,li2008built}, while others found no significant correlation~\cite{simmons2005choice, jeffery2006fast}. With the advent of mobile applications, researchers gained access to vast amounts of self-reported food choice data and could automatically track individuals' locations~\cite{allcott2019food}. For instance, large-scale dietary tracking data revealed an association between residential communities with greater healthy food accessibility and better dietary quality~\cite{althoff2022large}. However, as more food consumption occurs far from residential areas, focusing solely on residential environments may not fully capture individuals' exposure to food environments. Recent studies have used mobility data to examine urban residents' exposure to food environments, finding that higher exposure to fast food outlets in mobile environments increased the likelihood of visiting fast food restaurants~\cite{garcia2024effect}. Building on the concept of linking food environments with food choice behaviors, this study aims to extend previous knowledge by clearly defining the online food environment and quantifying its impact on online food choices.

\subsubsection{Online Service Data and Health of Citizens}

Digital data collected from online services, including activity tracking, GPS traces, and social media posts, have been widely linked to the health outcomes of urban residents in existing research~\cite{lin2020healthwalks, zhang2021passive, zhang2021quantifying, xu2025using}. Large-scale physical mobility data collected via smartphone applications has been used to reveal the discrepancy in urban residents' health status~\cite{chen2022strategic, zhang2024counterfactual, zhang2025metacity}. For example, such discrepancy significantly decreases as city walkability improves~\cite{althoff2017large}. A study based on mobile GPS data found that while large cities offer more recreational and social options, they tend to direct individuals to locations frequented by people of similar socioeconomic status, increasing the segregation of social activities and leading to new public health risks, such as lack of physical activity and imbalanced diets~\cite{nilforoshan2023human, chen2023getting, fan2025invisible}. Social media data, spanning multiple modalities, are also emerging as promising proxies for public health indicators. Posts on X (formerly Twitter) have been used to predict county-level public health metrics such as obesity and diabetes rates~\cite{culotta2014estimating, abbar2015you}. Food-related photos on Instagram, combined with location information, can accurately reflect offline food deserts~\cite{de2016characterizing}. YouTube videos have been applied to track dietary changes during the COVID-19 pandemic~\cite{mejova2023comfort}. Additionally, online recipe platforms are being used to nudge healthier food adoption by promoting specific healthy recipes~\cite{jesse2021digital, chelmis2023recipe}. Similarly, data generated by O2O food delivery platforms hold the potential to reflect the health-related information of urban residents.

\section{Data Description}

In this study, we mainly use three categories of data, \textit{i.e.}, O2O food delivery restaurant data, O2O food delivery consumption order data, and demographic data.

\subsubsection{O2O Food Delivery Restaurant Data}
We collect information on O2O food delivery restaurants from one of China's largest O2O food delivery platforms. For each restaurant, the platform provides its name, category, and service range. Each restaurant is assigned to a single, specific category. To assess the healthiness of the restaurants, we classify those within the ``fast food and snacks'', ``western fast food'', and ``fried food'' categories, as these are often regarded as having low nutritional value and may contribute to health issues such as obesity~\cite{hu2016higher}. All other categories are considered non-fast food restaurants. On average, 18.77\% of restaurants are categorized as fast food restaurant. 

\begin{figure}[t]
\centering
\includegraphics[width=\columnwidth]{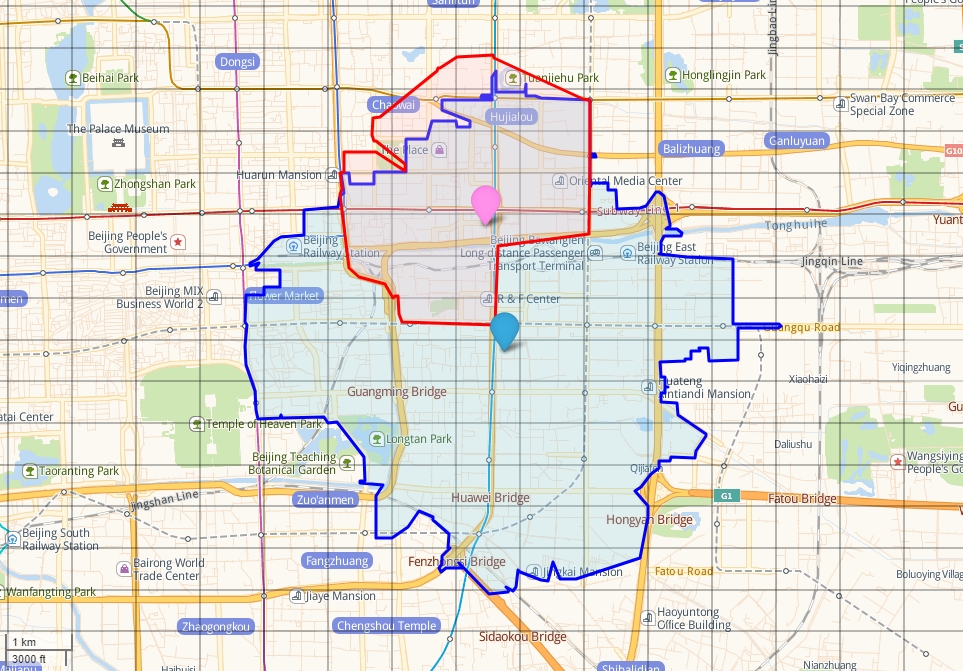}
\caption{Examples of service ranges for O2O food delivery restaurants.}
\label{fig:service_range}
\end{figure}

\begin{figure}[htbp]
\centering
\includegraphics[width=\columnwidth]{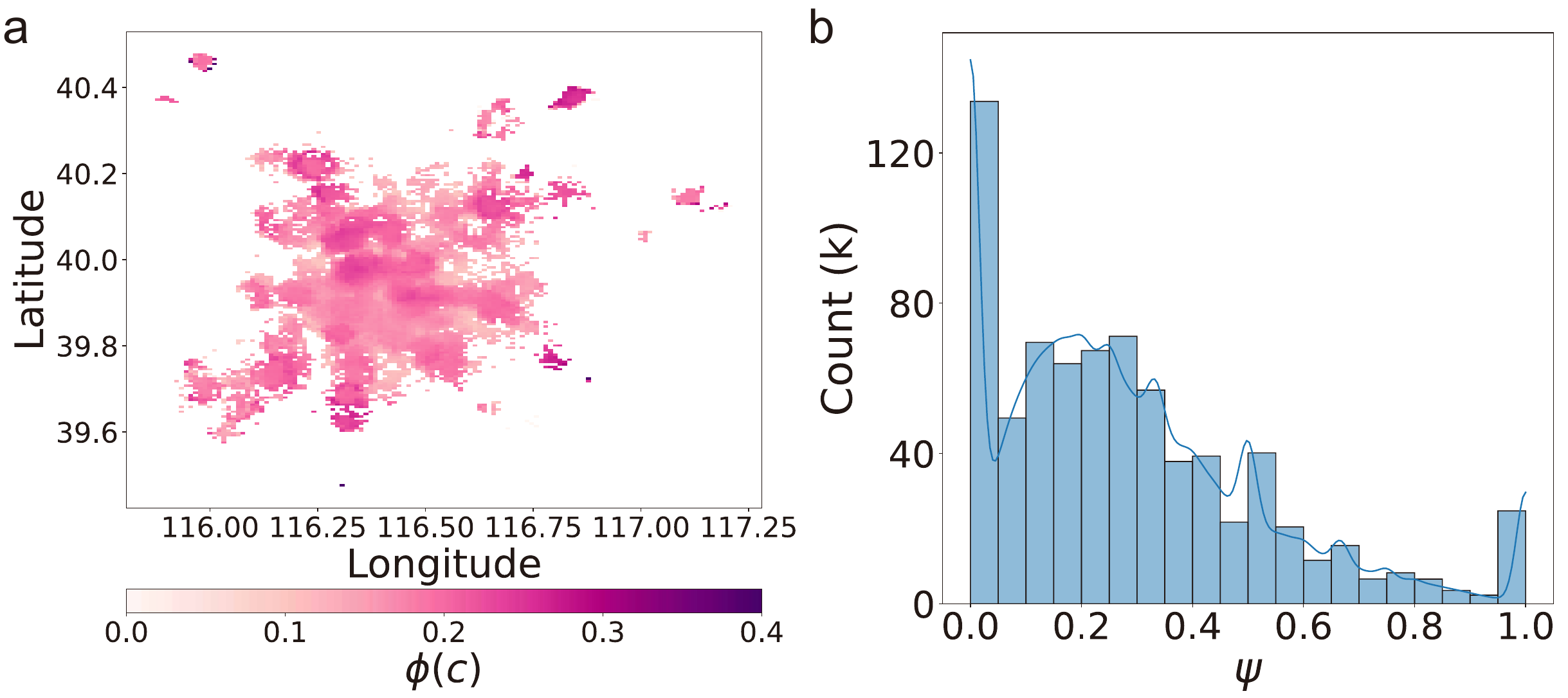}
\caption{Distributions of online food environment and preferences. (a) The spatial distribution of regions' ratio of accessible fast food restaurant $\phi(c)$ in Beijing. (b) The distribution of users' fast food preference $\psi$.}
\label{fig:data2}
\end{figure}

The service range of each O2O food delivery restaurant indicates the urban areas that can access its delivery services. As depicted in Figure~\ref{fig:service_range}, the restaurant marked in blue has a broader service range than the one marked in pink. Utilizing this data, we can identify O2O food delivery restaurants accessible to each urban area. We divide the city of Beijing into Geohash-6 ``regions'' (approximately 1.2$\times$0.6 kilometers), represented as rectangles in Figure~\ref{fig:data2}(a). For each region $c$, the ``unhealthiness'' of its corresponding online food environment $\phi(c)$ is defined as the ratio of the number of accessible fast food restaurants to the total number of accessible restaurants. Figure~\ref{fig:data2}(a) illustrates the spatial distribution of $\phi(c)$ across 4,921 regions in Beijing. 

\subsubsection{O2O Food Delivery Order Data}
To depict the healthy food choices of O2O food delivery users, we further collect O2O food delivery order data that record the consumption behaviors of platform users. Specifically, we gather a \textit{city-level} order number dataset and an \textit{individual-level} order record dataset from the same platform. 

The city-level dataset records the number of daily O2O food delivery orders, including fast food orders, across 283 Chinese cities in 2018 and 2024, respectively. For the individual-level dataset, we randomly sample 850,000 platform users in Beijing and collect their order records from January 2017 to May 2018. Additionally, we gathered order records from 20,000 users in Beijing, 10,000 users in Chengdu (a second-tier city in southwestern China), and 10,000 users in Xiamen (a third-tier city in southeastern China) for the period from January 2023 to December 2024. Each entry includes the user ID, order time, user location (at the Geohash-6 region level), expenditure amount, and the corresponding O2O food delivery restaurant. For the largest Beijing dataset from 2018, each user placed 50.3 orders on average. Based on these detailed order records, we calculate the fast food preference $\psi$ of each user, defined as the ratio of fast food orders to the total number of orders. Figure~\ref{fig:data2}(b) illustrates the distribution of $\psi$ on the sampled population. On average, 27.98\% of O2O food delivery orders are placed at fast food restaurants. 

\subsubsection{Demographic Data}

\begin{figure}[htbp]
\centering
\includegraphics[width=\columnwidth]{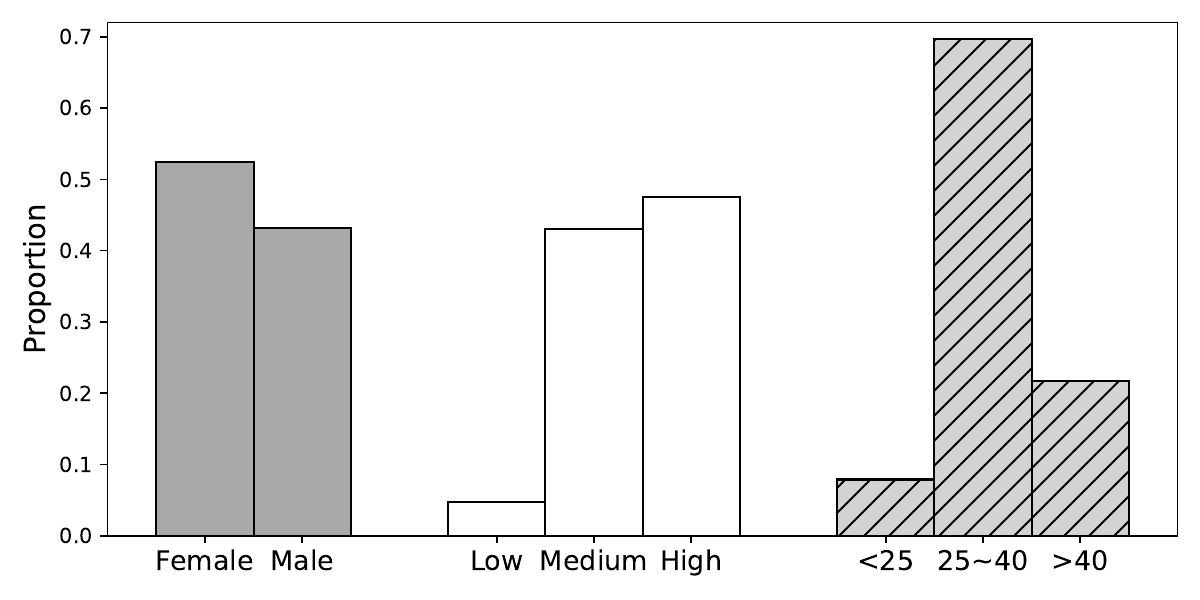}
\caption{Demographic distributions of sampled users.}
\label{fig:demographic}
\end{figure}

To establish a connection between healthy food choices and user attributes, we further obtain demographic data from the platform, which includes gender, income level, and age, as inferred by the platform's machine learning algorithms from user-provided profiles and user behaviors. We filter out users whose demographic profiles have a confidence level below 95\%. To ensure user privacy, income levels are categorized into three groups: ``high-income'', ``medium-income'' and ``low-income''. Similarly, age groups are segmented into ``below 25'', ``25 to 40'', and ``over 40''. The demographic distributions of the sampled population are depicted in Figure~\ref{fig:demographic}.

It is important to note that all the above datasets are processed under stringent user privacy protection protocols. Specifically, users provide consent for the access and use of their order behaviors and demographic information via a confidentiality agreement with the food delivery platform. The datasets are anonymized by the platform, ensuring that no actual user IDs are accessible during processing. To safeguard against online privacy breaches, the datasets are stored offline, and access is strictly limited to authorized researchers only. These precautions ensure that our analysis upholds high standards of user privacy protection.

\section{Methods and Results}

\subsection{Macro-level Health Discrepancy of O2O Food Delivery (RQ1)}

\subsubsection{Urban Scaling Laws of O2O Food Delivery}

We first demonstrate the macroscopic city-level health discrepancy of O2O food delivery. Urban services and economic activities often follow a scaling law, expressed as $Y_i=Y_0N_i^\alpha$~\cite{bettencourt2021introduction, xu2021emergence,ribeiro2023mathematical}, where $Y_i$ represents a measure of socioeconomic activity or resources, and $N_i$ denotes the city $i$'s population. An exponent $\alpha<1$ indicates sublinear growth of $Y$ relative to the city population, typically associated with urban infrastructure such as gas stations or road length, where the per-capita infrastructure decreases in larger cities~\cite{um2009scaling}. In contrast, socioeconomic factors like GDP or crime rates often exhibit $\alpha>1$, signifying superlinear growth, where per-capita values increase more rapidly in larger cities due to intensified social interactions~\cite{bettencourt2013origins, succar2024urban}. 

\begin{figure}[htbp]
\centering
\includegraphics[width=\columnwidth]{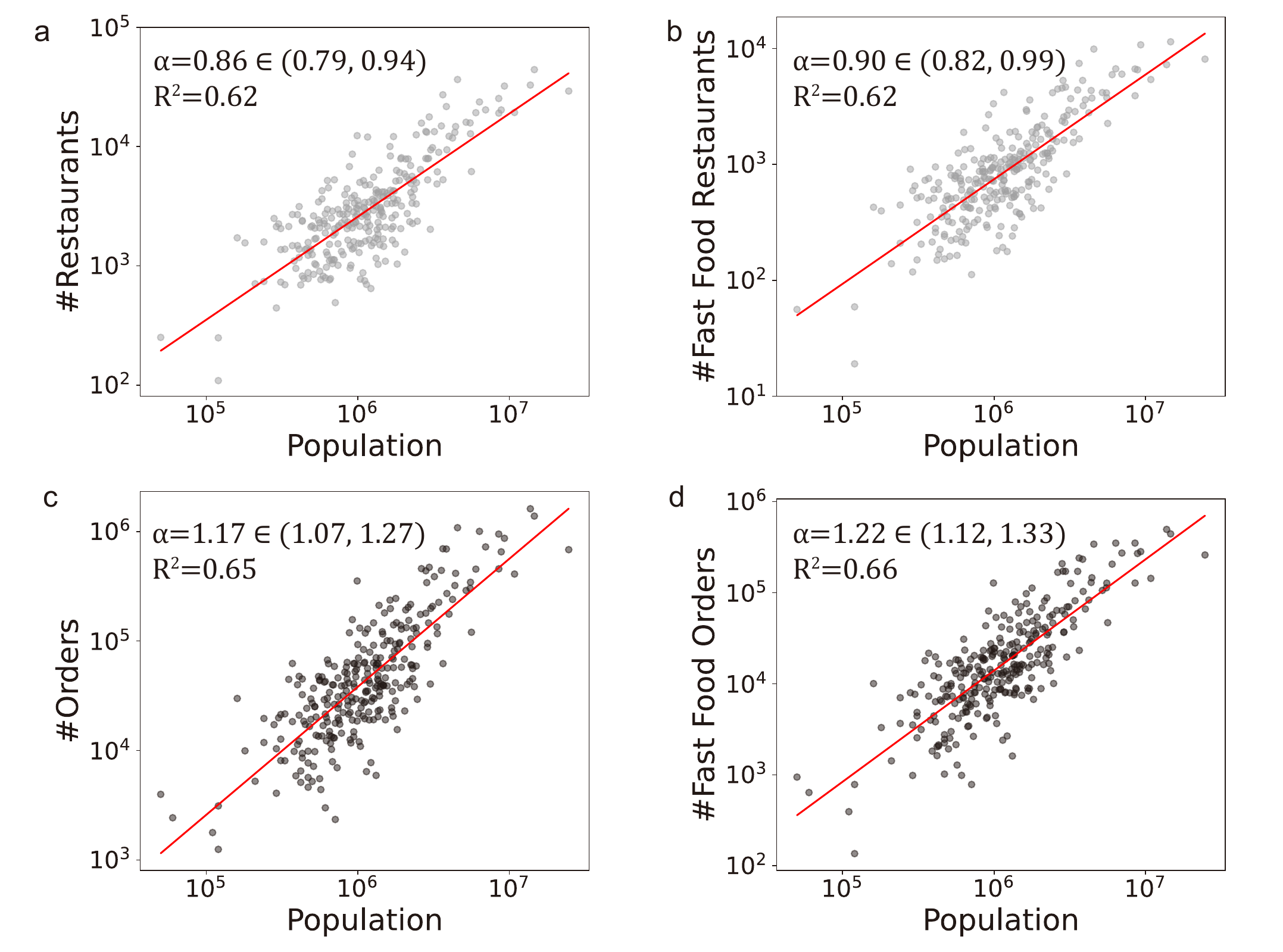}
\caption{Scaling laws of O2O food delivery restaurants (a), fast food restaurants (b), O2O food delivery orders (c), and fast food orders (d) in Chinese cities in 2018. The axes are in logarithmic scales.}
\label{fig:scaling}
\end{figure}

We collect city population from China City Statistical Yearbook\footnote{https://www.stats.gov.cn/sj/ndsj/2019/indexch.htm} and analyze the scaling law of O2O food delivery. As an innovative form of urban services, O2O food delivery in Chinese cities also exhibits scaling behavior. The fitted lines in Figure~\ref{fig:scaling}(a) and (c) indicate that the scaling exponent $\alpha$ for the number of O2O food delivery restaurants and the number of O2O food delivery orders are 0.86$\pm$0.07 and 1.17$\pm$0.10, respectively. This corresponds to the typical sublinear and superlinear growth patterns of urban infrastructure and economic activities, with $\beta=\frac{5}{6}$ and $\frac{7}{6}$, reflecting higher infrastructure utilization efficiency in large cities~\cite{bettencourt2013origins}. Similarly, these patterns are also observed in the 2024 dataset, as depicted in Figure~\ref{fig:scaling_new}.

Furthermore, the scaling exponents for fast food restaurants and orders in 2018 are significantly higher than those for all food categories ($P<0.05$, two-sided t-test), as shown in Figure~\ref{fig:scaling}(b) and (d). Specifically, doubling the city population increases the proportion of fast food restaurants by 4.0\% and the proportion of fast food orders by 5.4\%. This difference suggests a health discrepancy across cities, with larger cities demonstrating a greater reliance on fast food restaurants through O2O food delivery~\cite{bettencourt2007growth}. However, this trend is not observed in 2024 (Figure~\ref{fig:scaling_new}(b) and (d)), where the scaling coefficients for all food categories and fast food restaurants show no significant differences, indicating a shift toward healthier dining habits among metropolitan residents post-pandemic. 

\subsubsection{Health Discrepancy in Online Food Choices}

\begin{figure}[htbp]
\centering
\includegraphics[width=\columnwidth]{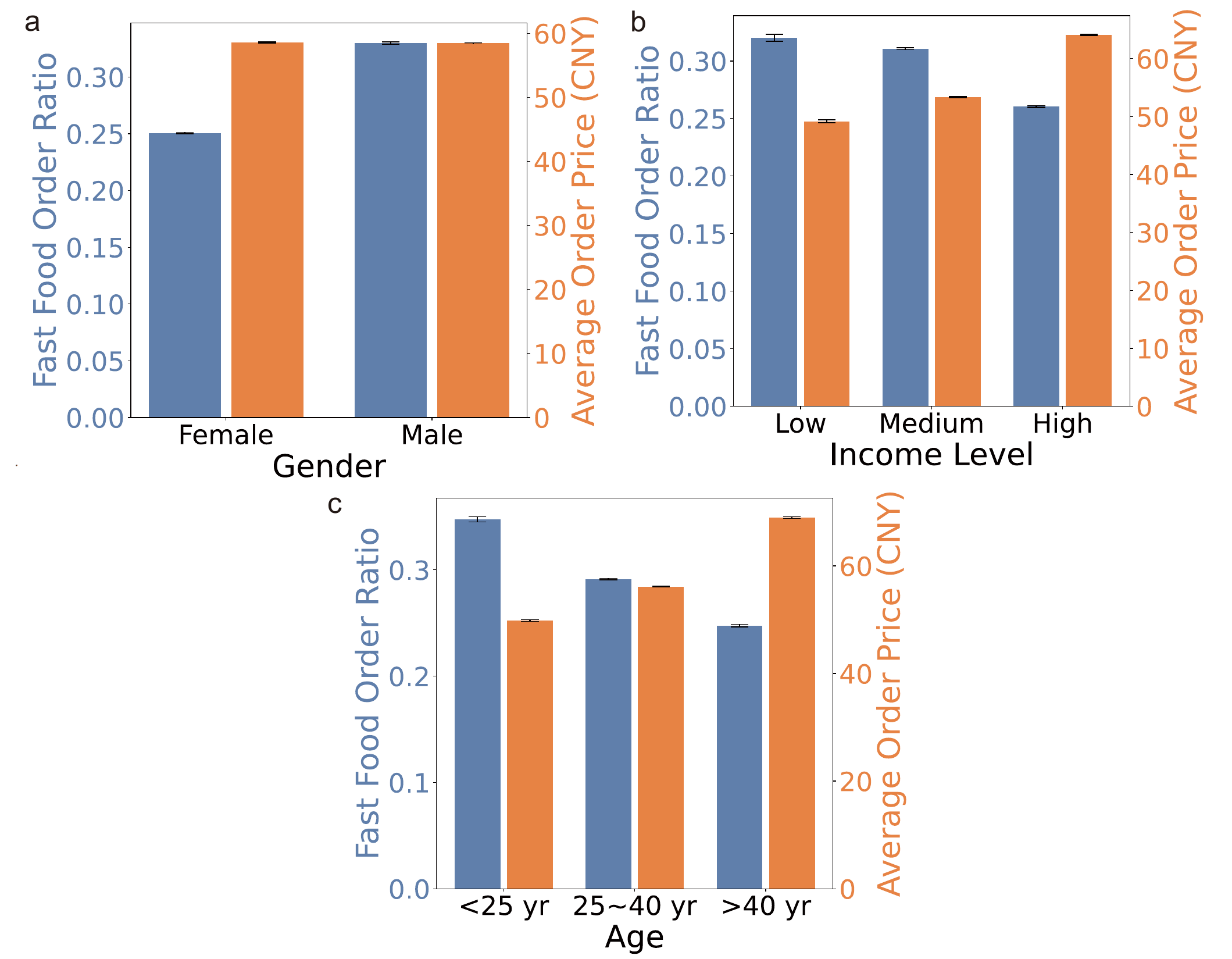}
\caption{The average fast food order ratio and order price across different gender (a), income level (b), and age (c) groups in Beijing. Error bars represent 95\% confidence intervals.}
\label{fig:ineq}
\end{figure}

The observed health discrepancies across cities of different sizes may be related to variations in the composition of urban populations. To test this assumption, we further investigate the relationships between user demographics and their behaviors on O2O food delivery platforms. We focus on three demographic attributes -- gender, income level, and age -- and compare the average proportion of fast food orders and the average order price for each group. The values and corresponding 95\% confidence intervals are shown in Figure~\ref{fig:ineq}.

The proportion of fast food orders among male users (33.0\%) is higher than that of female users (25.0\%) in 2018, while the average price of food delivery orders is nearly identical for males (58.4 CNY) and females (58.6 CNY). This indicates that male users have a significant stronger preference for fast food delivery compared to female users ($P<0.001$, two-sided t-test). In terms of income level, the average order price increases with higher income, while the proportion of fast food orders decreases. Users of highest income level only have 26.0\% fast food orders, while the ratio of the lowest income level reaches 32.0\%. This suggests that higher-income groups tend to opt for more expensive, healthy food delivery options, whereas lower-income groups prefer cheaper, less healthy food options ($P<0.001$, two-sided t-test). Similarly, the average price of food delivery orders rises with age, while the proportion of fast food orders declines from 34.8\% among users under 25 years to 24.8\% for users over 40 years, indicating that younger people are more inclined to choose fast food, while older individuals tend to prefer higher-priced alternatives ($P<0.001$, two-sided t-test).

This overall health discrepancy remains robust across three Chinese cities in 2024, as shown in Figure~\ref{fig:ineq_24}. This consistent trend highlights the heterogeneous healthy food choices of O2O food delivery platform users, which may be influenced by factors such as service pricing~\cite{zhang2025urban}, the online food environment, user preferences, and attention to healthy eating.

\subsection{Differences Between Offline and Online Food Environments (RQ2)}

O2O food delivery services extend the service range of local restaurants, though their impact on real-world food swamps remains unclear. In this section, we apply statistical learning algorithms to investigate the spatial clustering of offline and online food environments and compare their unhealthiness.

\begin{figure}[htbp]
\centering
\includegraphics[width=\columnwidth]{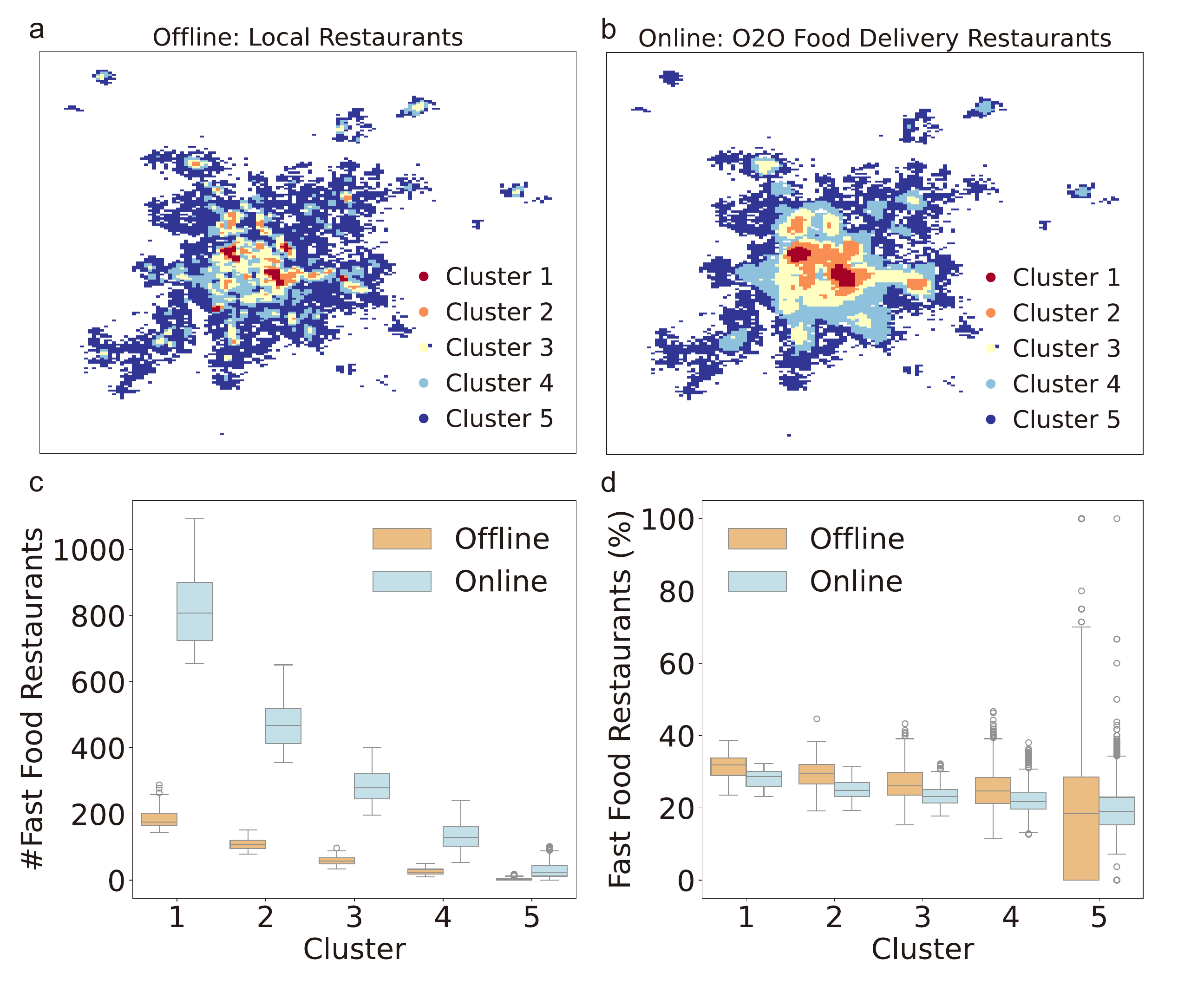}
\caption{The comparison of offline and online food swamps. (a) Spatial clustering based on regions' local restaurants. (b) Spatial clustering based on regions' accessible O2O food delivery restaurants. (c) Differences in the number of accessible fast food restaurants across clusters. (d) Differences in the proportion of accessible fast food restaurant across clusters.}
\label{fig:cluster_comparison}
\end{figure}

We cluster regions (Geohash-6 grids) based on the accessibility of local (offline) and O2O food delivery (online) restaurants, with each region represented by a 104-dimensional vector reflecting the number of accessible restaurants across 104 categories. Local restaurants are defined as those located within 1 kilometer of the region. We employ the K-means clustering algorithm~\cite{lloyd1982least} and use the elbow method~\cite{thorndike1953belongs} to determine that the optimal number of clusters is 5. As shown in Figure~\ref{fig:cluster_comparison}(a) and (b), both offline and online restaurant accessibility exhibit distinct spatial clustering. We compare the number of accessible fast food restaurants across each cluster in Figure~\ref{fig:cluster_comparison}(c). The core areas of the city (clusters 1 and 2) have the highest availability of O2O food delivery, with cluster 1, representing two key business districts (Zhongguancun and Guomao), having the highest access. 

The number of accessible online restaurants is approximately three times greater than that of offline restaurants, highlighting the pivotal role of O2O food delivery services in expanding access to convenience food options. Figure~\ref{fig:cluster_comparison}(d) shows the proportion of fast food restaurants ($\phi$) in each cluster for both offline and online food environments. In regions near the city center, the proportion of fast food restaurants is higher. Notably, the proportion of fast food in the online food environment is slightly lower than in the offline environment (with regional averages of 21.2\% and 22.1\%, respectively), though the online food environment still contains a substantial proportion of fast food. Moreover, by comparing the red and orange clusters in Figure~\ref{fig:cluster_comparison}(a) and (b), we observe that O2O food delivery has expanded the size of clusters with the highest fast food count and proportion. This indicates that O2O delivery services may further exacerbate real-world food swamps, potentially creating a ``cyber food swamp''.

\begin{figure}[htbp]
\centering
\includegraphics[width=\columnwidth]{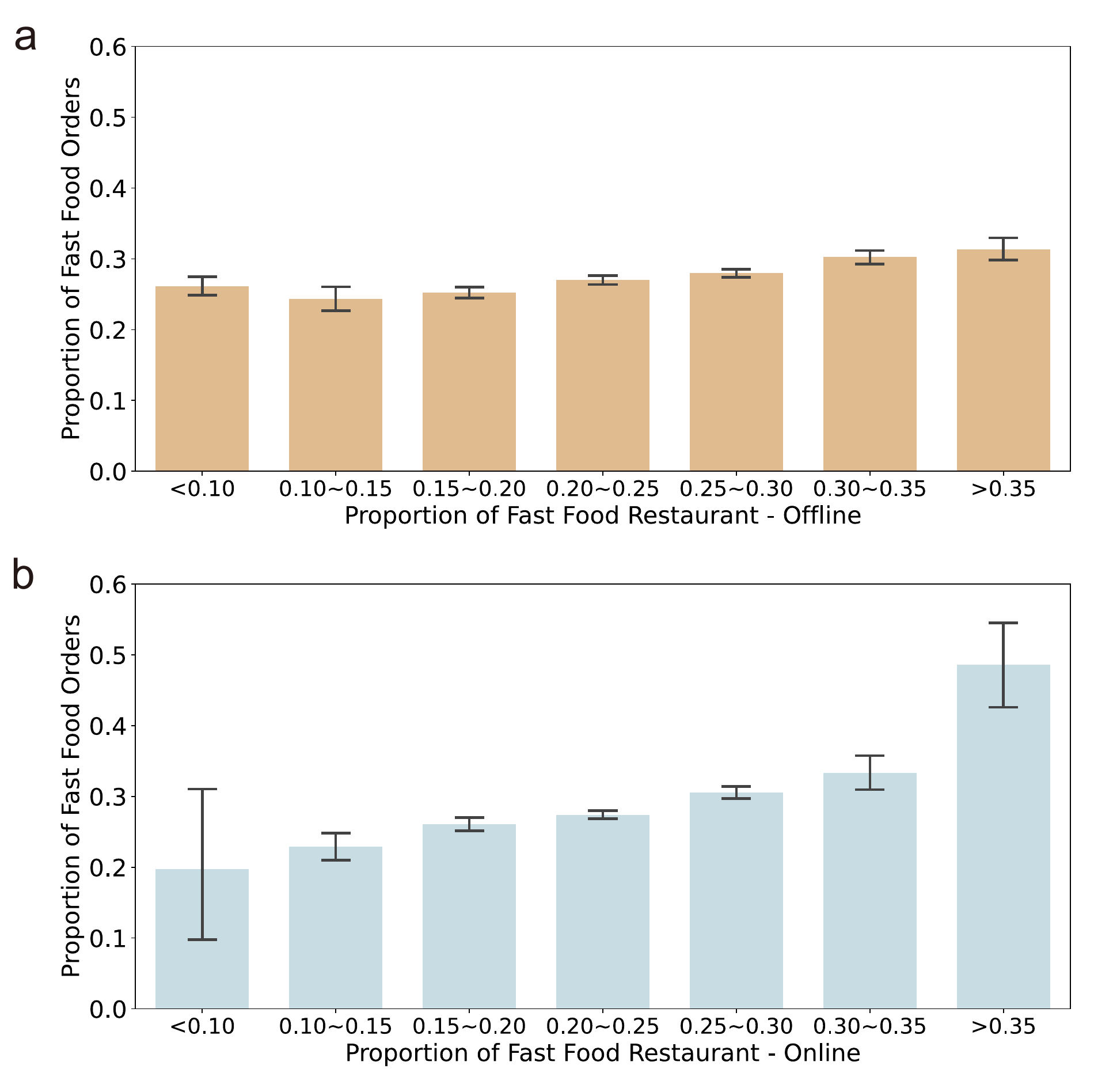}
\caption{The average choices on fast food orders of regions within different levels of offline (a) and online (b) food environment $\phi$, as represented by the proportion of accessible fast food restaurants. Error bars indicate 95\% confidence intervals.}
\label{fig:order_env}
\end{figure}

By comparing the unhealthiness $\phi$ of offline and online food environments with fast food preferences in regional O2O food delivery orders, we find that the correlation is stronger for online environments ($r = 0.204$) compared to offline environments ($r = 0.092$). Figure~\ref{fig:order_env} demonstrates that variations in online food environments are associated with more distinct differences in healthy food choices. In Figure~\ref{fig:order_env}(b), regions with lowest $\phi$ exhibit significant lower proportion of fast food orders compared with regions with $\phi>0.25$ ($P<$0.05, two-sided t-test). These findings suggest notable differences between online and traditional offline food environments. The stronger relationship between the online food environment and healthy food choices underscores the need for further investigation into their causal connections.

\subsection{Impact of Online Food Environment on Healthy Food Choice (RQ3)}

Given the diversity of online food environments within urban contexts and the heterogeneous healthy food choices exhibited by users on O2O food delivery platforms, it is essential to understand the potential impact of these environments on food choices. Previous studies have demonstrated that increased exposure to fast food restaurants can contribute to higher obesity rates~\cite{cooksey2017food} and increased visitation~\cite{garcia2024effect}. As O2O food delivery significantly expand the range of available restaurant options, they hold substantial potential to influence users' healthy food choices.

To accurately quantify the relationship between the environment and healthy food choices, it is essential to dissect the influence of users' personal preferences from the impact of the environment. For instance, individuals may choose to order food from a fast food restaurant either due to a personal preference for fast food or because the delivery environment is saturated with fast food options. Building on this notion, we adopt the analytic framework in Garc{\'\i}a Bulle Bueno \textit{et al.}~\cite{garcia2024effect}. We first conduct an individual-level regression analysis on the determinants of O2O food delivery orders, followed by a quasi-natural experiment to quantify the causal impacts of changes in online food environment on healthy food choices.

\subsubsection{Regression Analysis} 
\begin{figure}[t]
\centering
\includegraphics[width=\columnwidth]{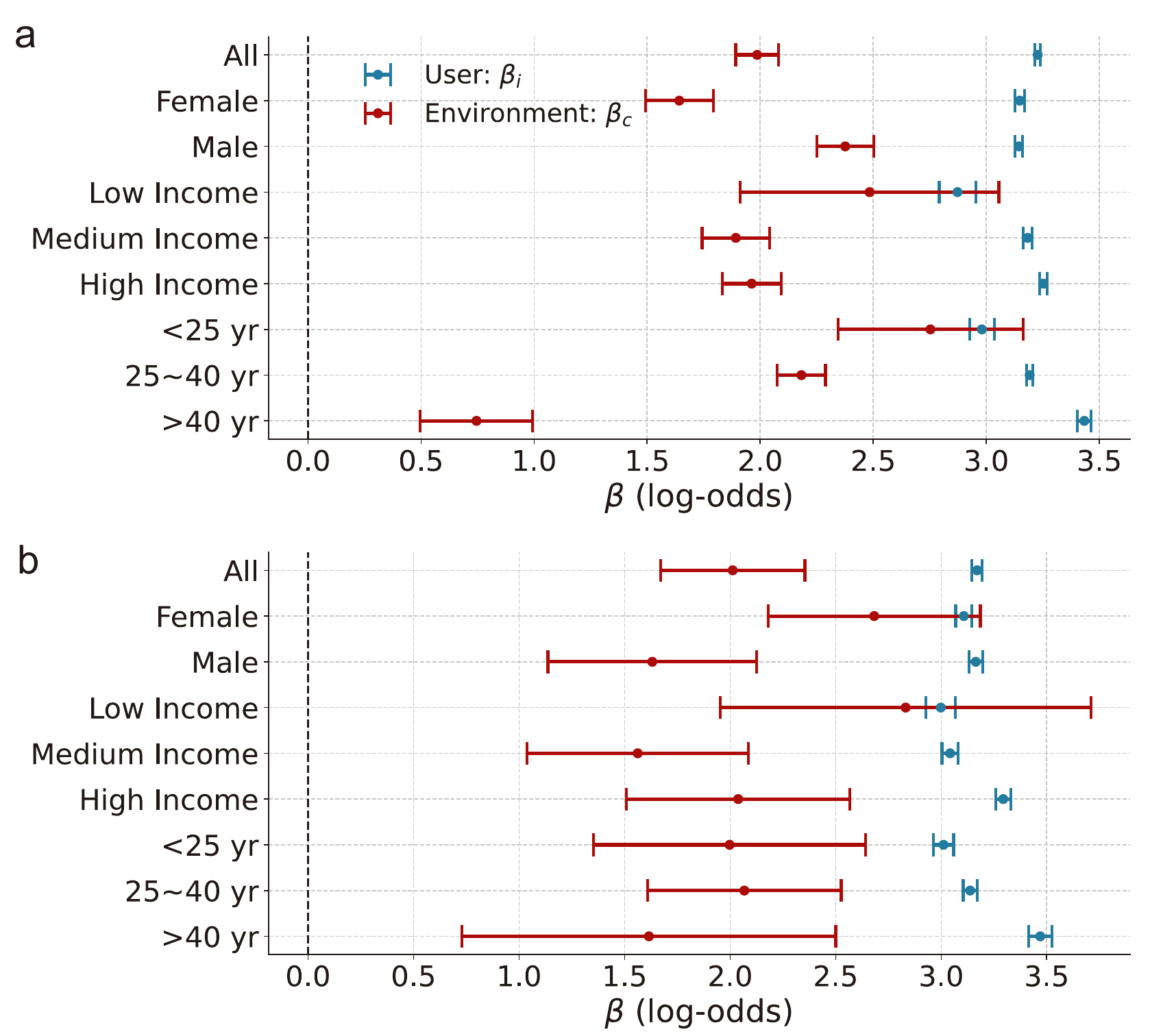}
\caption{Effect of user's historical preference ($\beta_i$) and online food environment ($\beta_c$) on the probability of placing a fast food delivery order over a non-fast food delivery order for different gender, income, and age groups in 2018 (a) and 2024 (b). Error bars represent 95\% confidence intervals for $\beta$'s.}
\label{fig:lr}
\end{figure}

For the online food order decision $y_{it}$ by user $i$ located in region $c_{it}$ at time $t$, we characterize the online food environment using $\phi(c_{it})$ (the proportion of accessible fast food restaurants), and user $i$'s fast food preference $\psi_i(t)$, as the proportion of fast food orders made by user $i$ within six months prior to time $t$. We focus on 91,879 users who have made at least 20 orders, extracting 3.86 million orders to fit a logistic regression model to estimate the effect of online food environment $\phi(c_{it})$ on the decision to choose fast food over non-fast food:

\begin{equation}
    \text{Pr}(y_{it}=1)=\text{logit}^{-1}(\beta_0+\delta_t+\beta_i \psi_i(t)+\beta_c\phi(c_{it})),
\end{equation}
where $y_{it}=1$ represents ordering fast food, $\text{logit}^{-1}(x)=\frac{e^x}{1+e^x}$ is the logistic function, $\beta_0$ is the fixed intercept term, and $\delta_t$ is a fixed effect term accounting for monthly variation. $\beta_i$ and $\beta_c$ characterize the separate effect of user preference and online food environment, respectively, as shown in Figure~\ref{fig:lr} and Figure~\ref{fig:lr_appendix}.

The regression model reveals that both online food environment and personal preference are significantly associated with fast food choice, with personal preference exhibiting a higher log-odd coefficient ($\beta_i=3.227\pm0.012$ in 2018, $\beta_i=3.170\pm0.024$ in 2024, $P<0.001$, two-sided t-test) than the online food environment ($\beta_c=1.987\pm0.095$ in 2018, $\beta_c=2.012\pm0.342$ in 2024, $P<0.001$, two-sided t-test). These results can be interpreted as follows: when a user has a 10\% higher historical preference for fast food, the odds of making a fast food order increase by $e^{\beta_i\times \text{0.1}}-1$=38.1\% for 2018 and 37.3\% for 2024, and when the online food environment consists of 10\% more fast food restaurants, the odds increase by $e^{\beta_c\times \text{0.1}}-1$=22.0\% for 2018 and 22.3\% for 2024. Similar patterns are observed in Chengdu and Xiamen. Although the influence of the online food environment is relatively minor, it still has a significant association with users' consumption behaviors, suggesting a potential negative impact of ``cyber food swamps'' on health outcomes.

We also observe heterogeneous effects of the online food environment on healthy food choices across different demographic groups. Figure~\ref{fig:lr} illustrates the fitted log-odds $\beta$'s for orders made by specific demographic groups. Across all groups, user preference consistently has a stronger influence than the online food environment, although the magnitude of $\beta_c$ varies between groups. Low-income individuals and users under the age of 25 are more influenced by the environment and less by personal preference compared to their counterparts. Notably, users over the age of 40 are the least influenced by the online food environment, with $\beta_c=0.745\pm0.249$ in 2018, meaning that the odds of ordering fast food increase by only 7.7\% when the online food environment contains 10\% more fast food restaurants. This pattern aligns with the disparity trend shown in Figure~\ref{fig:ineq}, where groups more prone to making less healthy food delivery choices are also more affected by the online food environment. This suggests that these users may be more susceptible to online platform's recommender system when ordering food~\cite{li2022exploratory, sukiennik2024uncovering}, rather than using the platform to place an order after already deciding what to eat. 

\subsubsection{Causal Impacts of Online Food Environment on Healthy Food Choice}

Although the regression analysis confirms an association between the online food environment and consumers' healthy food choices, our understanding of the causal impact of the environment on food choices remains limited, especially when the influence is gradual and non-immediate as users adapt to a new location. To investigate this causal impact, we follow the semi-causal framework introduced in Garc{\'\i}a Bulle Bueno \textit{et al.}~\cite{garcia2024effect} and design a quasi-natural experiment to examine the relationship between order context and decision-making. The approach involves identifying users who have experienced a significant shift in their food delivery context, represented by the Geohash-5 grid in which they are located, and quantifying how the change in the online food environment $\phi$ between two contexts affects their fast food order preference $\psi$. This setting simulates a natural experiment scenario where users randomly shift their context, allowing for the derivation of causal impacts.

\begin{figure}[htbp]
\centering
\includegraphics[width=\columnwidth]{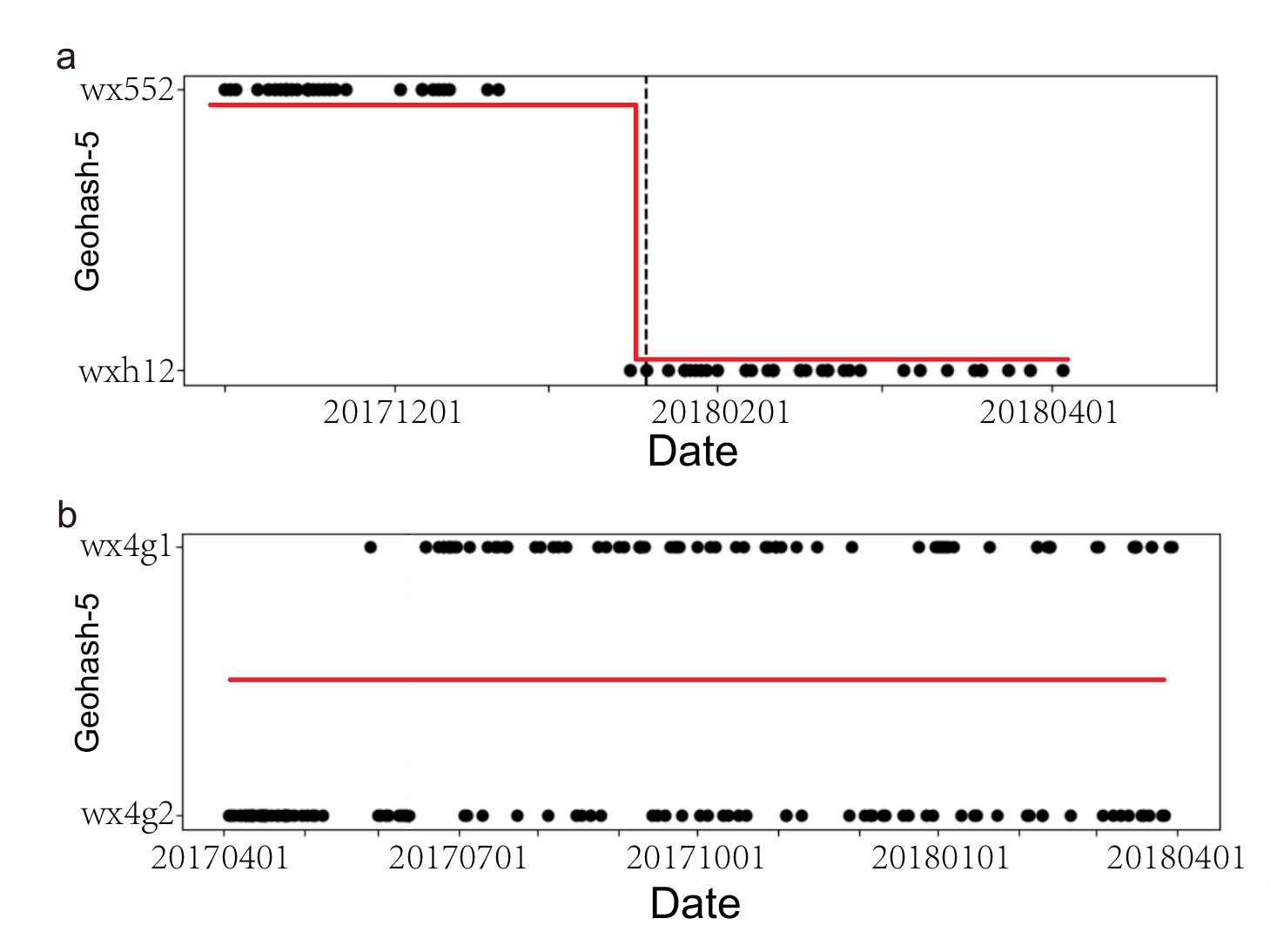}
\caption{Examples of detecting the change of food delivery order context, defined as the Geohash-5 grid where user make food delivery orders. (a) Significant context change detected in January 2018. (b) No significant context change detected.}
\label{fig:changepoint}
\end{figure}

To identify users who changed their preferred food delivery locations during the observation period, we extract the Geohash-5 grids (approximately 5$\times$5 kilometers) for each order and select users who placed more than five orders across two different grids, with each grid representing more than 30\% of their total orders. We then construct a binary time series for these users, capturing their order activity in the two grids, as shown in Figure~\ref{fig:changepoint}. We apply a mean change detection method~\cite{truong2020selective} to identify if and when substantial changes occurred in the time series, enforcing a minimum spatial shift of 2 kilometers and a minimum interval of 21 days between change points to avoid mis-detecting fluctuations. Figure~\ref{fig:changepoint}(a) illustrates a successful detection of a context change, where the user shifted from grid ``wx552'' to ``wxh12'' in January 2018. In contrast, the user in Figure~\ref{fig:changepoint}(b) uniformly used O2O food delivery services across two grids, with no significant change detected. Using this method, we identify 1,892 users who experienced a substantial change in their order contexts in 2018. On average, the spatial shifts in their order context are 11.66 kilometers, indicating that their online food delivery environment has completely changed, as supported by Figure~\ref{fig:service_range}.

\begin{figure}[t]
\centering
\includegraphics[width=\columnwidth]{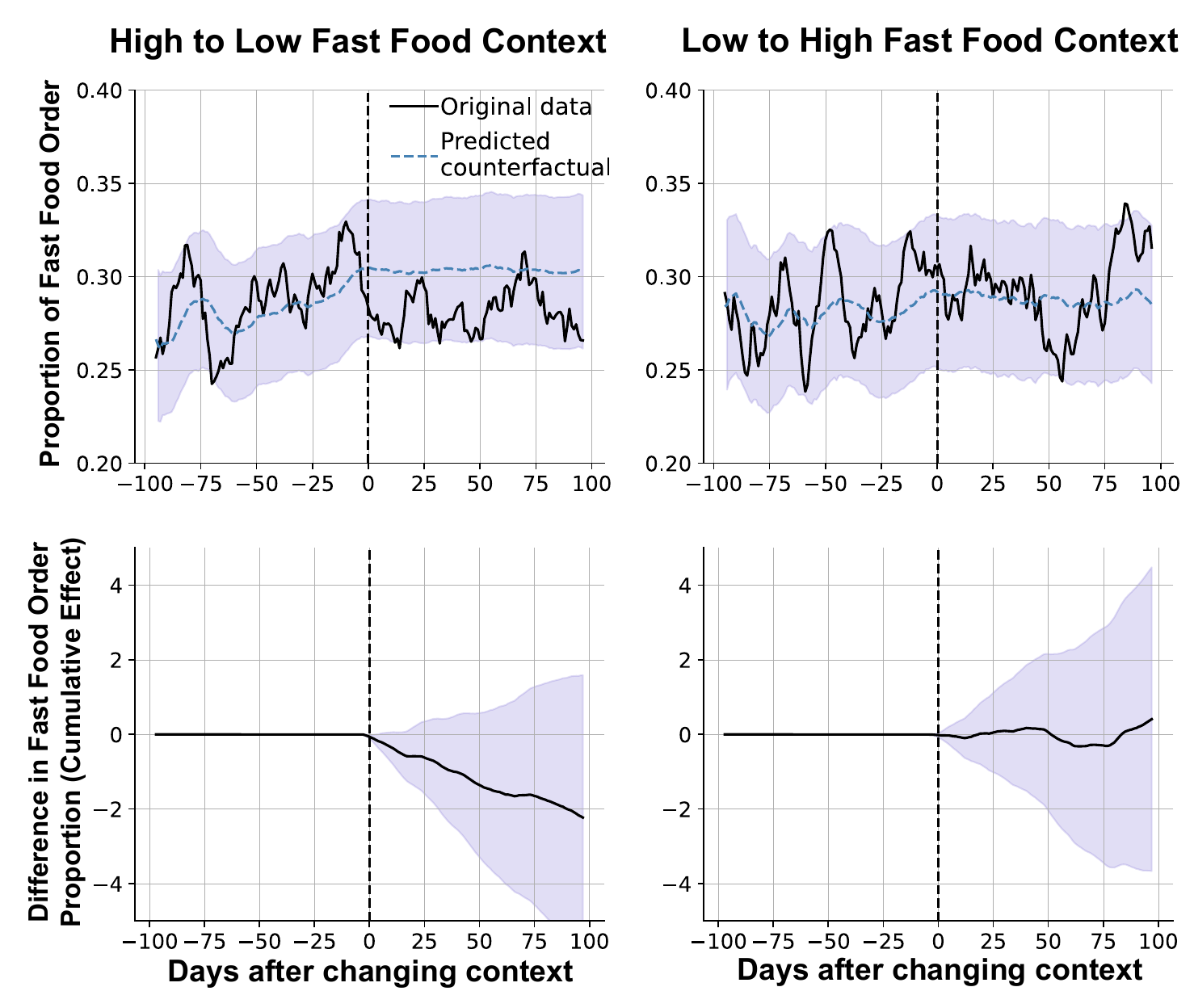}
\caption{Evolution of the proportion of fast food orders (top) and cumulative differences in proportion of fast food order (bottom) for user groups transitioning from a high fast food ratio online food environment to a low ratio context (left), and from low to high context (right) in Beijing, 2018. Shaded areas represent 50\% confidence intervals of predicted counterfactual and cumulative effects.}
\label{fig:intervention}
\end{figure}

Based on whether the user’s order context before and after the change is classified as a high fast food context ($\phi>$0.2) or a low fast food context ($\phi<$0.2), we divide users into four groups. Among them, 957 moved from a low to low fast food context, 481 from a high to high context, 208 from a low to high context, and the remaining 246 from a high to low context. To estimate the causal impact of changes in the online food environment on users' healthy food choices, we employed a method based on Bayesian Structural Time Series (BSTS)~\cite{brodersen2015inferring}. BSTS decomposes time series into multiple structural components to predict temporal behaviors. To assess the effect of shifting from a low to high fast food context, we take users moved from a low to low context as the control group, and those who moved from a low to high context as the experimental group. We compute the time series $\{\psi_t|-100<t<100\}$ for each group, where $\psi_t$ represents the proportion of fast food orders after $t$ days of changing context. BSTS then predicts the counterfactual outcome for the experimental group based on $\{\psi_t^{\text{Experiment}}|t<0\}$ by applying the temporal patterns of $\{\psi_t^{\text{Control}}\}$, as shown by the dashed blue lines in Figure~\ref{fig:intervention}. The counterfactual outcomes are then compared with the actual observations of $\{\psi_t^{\text{Experiment}}|t>0\}$ to derive the causal impact. A similar procedure is applied to estimate the effect of shifting from a high to low context, with users who moved from a high to high context as the control group.

\begin{figure}[htbp]
\centering
\includegraphics[width=\columnwidth]{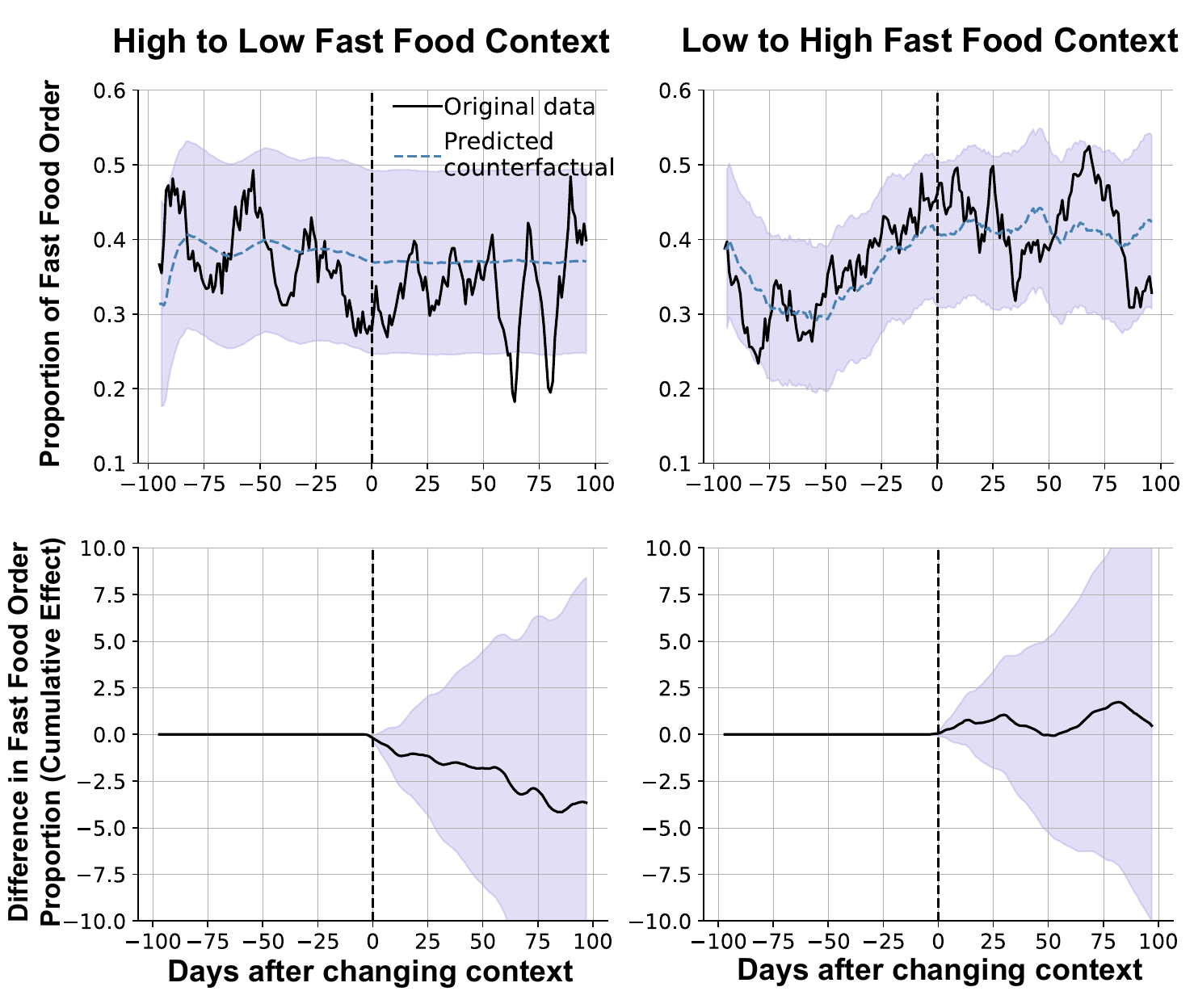}
\caption{Evolution of the proportion of fast food orders (top) and cumulative differences in proportion of fast food order (bottom) for user groups transitioning from a high fast food ratio online food environment to a low ratio context (left), and from low to high context (right) in Beijing, 2024. Shaded areas represent 50\% confidence intervals of predicted counterfactual and cumulative effects.}
\label{fig:intervention_rev}
\end{figure}

As shown in Figure~\ref{fig:intervention}, for users moving from a low to high fast food context, their preference for using O2O food delivery to order from fast food restaurants increases by 1.6\%, with a cumulative proportion of fast food order of 0.46 times higher than that of users who remained in a low fast food context. In contrast, users who moved from a high to low fast food context experience a 7.6\% decline in their proportion of fast food orders, with a cumulative effect 2.33 times lower than users who stayed in a high fast food context after 100 days. Similar trends were observed in 502 users who shifted their online food environment in 2024, as depicted in Figure~\ref{fig:intervention_rev}. Specifically, moving from a low to high fast food context leads to a 0.2\% increase in the proportion of fast food orders, while shifting from a high to low fast food context results in a 9.6\% decrease in the proportion of fast food orders. These results confirm that the online food environment affects users' healthy food choices, and this influence is not merely a short-term response to visiting a new location but has a lasting impact. Moreover, the negative health impact of moving to high fast food context is less substantial than the positive health impact of moving to a low fast food context. This finding suggests that O2O food delivery platforms could mitigate the negative health effects of ``cyber food swamps'' by recommending much healthier restaurants to users, even in environments with a high concentration of fast food options.

\section{Discussion}

Our findings substantiate the transformative role of O2O food delivery industry in reshaping urban residents' healthy food choices, providing valuable insights for various stakeholders. For O2O food delivery platforms, enhancing recommendation algorithms to guide users towards healthier options is critical. This can be accomplished through personalized search features that prioritize healthier restaurants or meals~\cite{vanderlee2023recommended}, along with clear nutritional labeling, such as calorie counts or warnings about high sugar or fat content~\cite{jiang2019market2dish, greenthal2023availability}, particularly targeting demographic groups such as males, low-income individuals, and younger users with a high preference for fast food~\cite{orfanos2009eating, adams2015prevalence}. Restaurants stakeholders can broaden their menus to include a wider variety of healthier choices, demonstrating their commitment to both food quality and public health~\cite{vanderlee2023recommended}. Policy makers must address the spatial clustering of cyber food swamps with targeted regulations. For example, they could provide financial incentives and subsidies to encourage O2O food delivery restaurants that offer healthier alternatives~\cite{shill2012government}. In addition, regulations should cap the concentration of fast food options in areas characterized by these swamps. Public health campaigns in schools and communities can significantly elevate awareness around healthy eating, a factor strongly associated with healthier dietary choices and improved health outcomes~\cite{iqbal2021health}. Moreover, stricter labeling regulations, modeled after initiatives like the Affordable Care Act in the United States~\cite{stein2010national}, are essential to ensuring transparent nutritional information for O2O food delivery items.

Our research also highlights the influence of online content in offline real-world social behaviors, aligning with previous studies on the web and social media~\cite{grinberg2013extracting, hu2015predicting, de2016characterizing, carpinelli2024quiet}. Meanwhile, O2O food delivery serve as both a substitute for and an expansion of traditional offline dining environments, creating potential trade-offs and interactions that reshape conventional food accessibility~\cite{li2022neighborhood}. It remains unclear whether O2O food delivery intensifies or mitigates food swamps. In future research, we aim to compare the offline and online dining behaviors of the same group of users to better understand the unique role of online platforms in shaping dietary habits, as distinct from traditional in-person dining experiences.

Our work has several limitations. First, we use whether an order is placed at a fast food restaurant as a proxy for healthy food choices. While fast food is generally associated with lower dietary quality, the nutritional content of items sold within the same restaurant can vary. Future research should assess the nutritional quality of foods purchased through O2O food delivery at the item level, such as sugar and fat intake, using more granular data. Additionally, the uncertainty in user demographic profiles could impact the validity of the observed differences across demographic groups. However, the platform reports that the inferred demographic profiles have an accuracy of over 90\%, and we filter out users with low confidence levels to ensure the reliability of our findings. Furthermore, other observations that do not involve comparisons across demographic groups remain robust and unaffected by this potential uncertainty.

\section{Acknowledgements}

This work is supported by the National Key Research and Development Program of China under grant No. 2024YFC3307600. This work is supported by the National Natural Science Foundation of China under grant No. 23IAA02114 and 62472241. This work is supported in part by Tsinghua University-Toyota Research Center.

\bibliography{aaai25}

\section{Paper Checklist}

\begin{enumerate}

\item For most authors...
\begin{enumerate}
    \item  Would answering this research question advance science without violating social contracts, such as violating privacy norms, perpetuating unfair profiling, exacerbating the socio-economic divide, or implying disrespect to societies or cultures?
    \answerYes{Yes}
  \item Do your main claims in the abstract and introduction accurately reflect the paper's contributions and scope?
    \answerYes{Yes}
   \item Do you clarify how the proposed methodological approach is appropriate for the claims made? 
    \answerYes{Yes}
   \item Do you clarify what are possible artifacts in the data used, given population-specific distributions?
    \answerNA{NA}
  \item Did you describe the limitations of your work?
    \answerYes{Yes}
  \item Did you discuss any potential negative societal impacts of your work?
    \answerNA{NA}
   \item Did you discuss any potential misuse of your work?
    \answerNA{NA}
   \item Did you describe steps taken to prevent or mitigate potential negative outcomes of the research, such as data and model documentation, data anonymization, responsible release, access control, and the reproducibility of findings?
    \answerYes{Yes}
  \item Have you read the ethics review guidelines and ensured that your paper conforms to them?
    \answerYes{Yes}
\end{enumerate}

\item Additionally, if your study involves hypotheses testing...
\begin{enumerate}
  \item Did you clearly state the assumptions underlying all theoretical results?
    \answerNA{NA}
  \item Have you provided justifications for all theoretical results?
    \answerNA{NA}
  \item Did you discuss competing hypotheses or theories that might challenge or complement your theoretical results?
    \answerNA{NA}
  \item Have you considered alternative mechanisms or explanations that might account for the same outcomes observed in your study?
    \answerNA{NA}
  \item Did you address potential biases or limitations in your theoretical framework?
    \answerNA{NA}
  \item Have you related your theoretical results to the existing literature in social science?
    \answerNA{NA}
  \item Did you discuss the implications of your theoretical results for policy, practice, or further research in the social science domain?
    \answerNA{NA}
\end{enumerate}

\item Additionally, if you are including theoretical proofs...
\begin{enumerate}
  \item Did you state the full set of assumptions of all theoretical results?
    \answerNA{NA}
	\item Did you include complete proofs of all theoretical results?
    \answerNA{NA}
\end{enumerate}

\item Additionally, if you ran machine learning experiments...
\begin{enumerate}
  \item Did you include the code, data, and instructions needed to reproduce the main experimental results (either in the supplemental material or as a URL)?
   \answerNA{NA}
  \item Did you specify all the training details (e.g., data splits, hyperparameters, how they were chosen)?
    \answerNA{NA}
     \item Did you report error bars (e.g., with respect to the random seed after running experiments multiple times)?
    \answerNA{NA}
	\item Did you include the total amount of compute and the type of resources used (e.g., type of GPUs, internal cluster, or cloud provider)?
   \answerNA{NA}
     \item Do you justify how the proposed evaluation is sufficient and appropriate to the claims made? 
    \answerNA{NA}
     \item Do you discuss what is ``the cost`` of misclassification and fault (in)tolerance?
    \answerNA{NA}
  
\end{enumerate}

\item Additionally, if you are using existing assets (e.g., code, data, models) or curating/releasing new assets, \textbf{without compromising anonymity}...
\begin{enumerate}
  \item If your work uses existing assets, did you cite the creators?
    \answerNA{NA}
  \item Did you mention the license of the assets?
    \answerNA{NA}
  \item Did you include any new assets in the supplemental material or as a URL?
    \answerNA{NA}
  \item Did you discuss whether and how consent was obtained from people whose data you're using/curating?
    \answerNA{NA}
  \item Did you discuss whether the data you are using/curating contains personally identifiable information or offensive content?
    \answerNA{NA}
\item If you are curating or releasing new datasets, did you discuss how you intend to make your datasets FAIR?
\answerNA{NA}
\item If you are curating or releasing new datasets, did you create a Datasheet for the Dataset? 
\answerNA{NA}
\end{enumerate}

\item Additionally, if you used crowdsourcing or conducted research with human subjects, \textbf{without compromising anonymity}...
\begin{enumerate}
  \item Did you include the full text of instructions given to participants and screenshots?
    \answerNA{NA}
  \item Did you describe any potential participant risks, with mentions of Institutional Review Board (IRB) approvals?
    \answerNA{NA}
  \item Did you include the estimated hourly wage paid to participants and the total amount spent on participant compensation?
    \answerNA{NA}
   \item Did you discuss how data is stored, shared, and deidentified?
   \answerNA{NA}
\end{enumerate}

\end{enumerate}

\section{Ethical Statement}

All datasets collected from the O2O food delivery platform were processed under strict privacy protection protocols. User IDs were anonymized, and demographic attributes were injected with noise and coarsely categorized into no more than three levels for age and income. All geographic coordinates, including the locations of the O2O food delivery restaurants and delivery destinations, were recorded at the Geohash-6 level (approximately 1.2$\times$0.9 kilometers) to prevent the disclosure of individual locations. All analyses were conducted on offline storage with strict data access regulations. These precautions ensure that our analysis maintains high standards of user privacy protection. In accordance with data confidentiality agreements, we are unable to share the raw data used in this study.

\appendix
\section{Appendix}

\begin{figure}[b!]
\centering
\includegraphics[width=0.81\columnwidth]{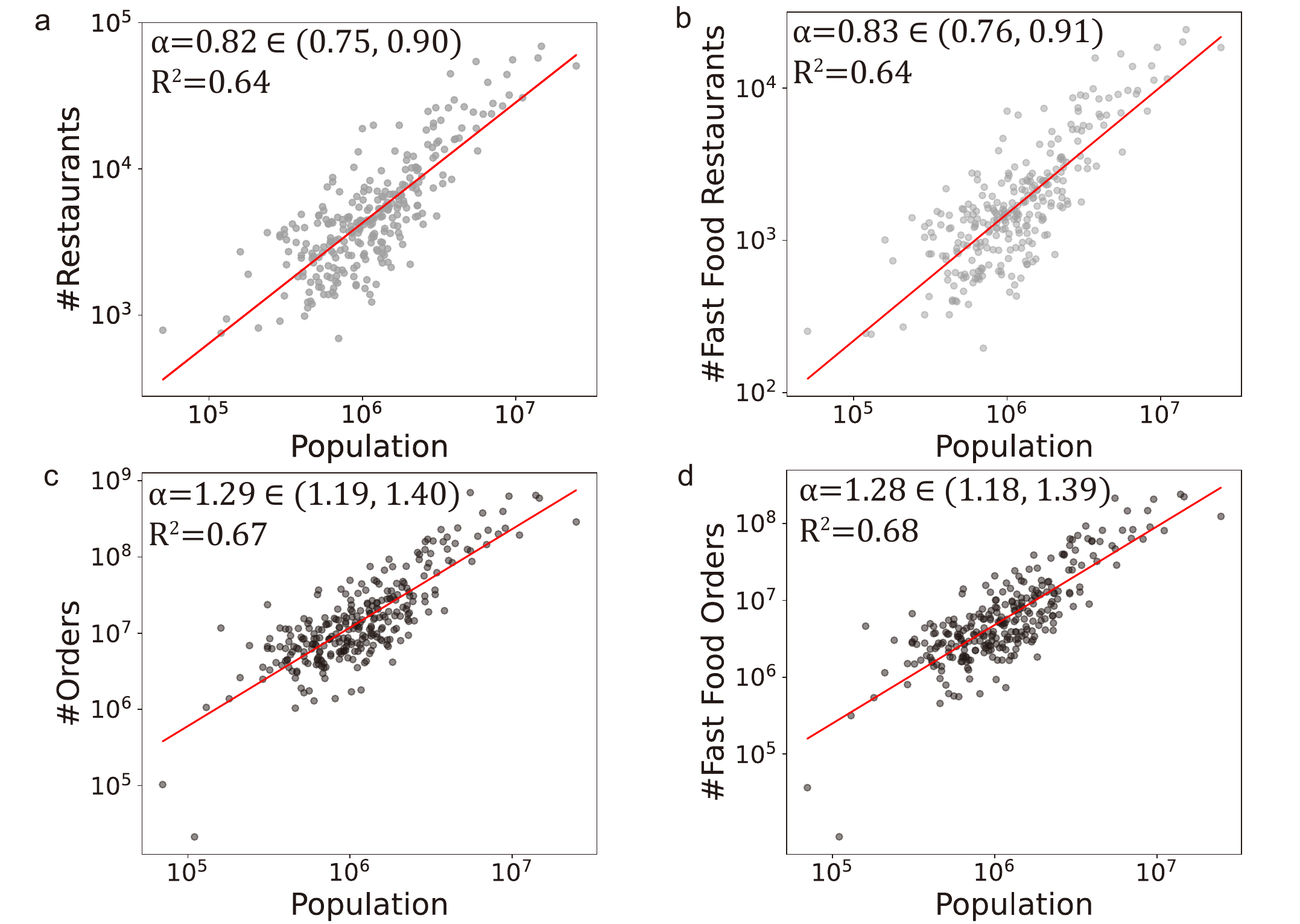}
\caption{Scaling laws of O2O food delivery restaurants (a), fast food restaurants (b), O2O food delivery orders (c), and fast food orders (d) in Chinese cities in 2024.}
\label{fig:scaling_new}
\end{figure}

\begin{figure}[htbp]
\centering
\includegraphics[width=0.81\columnwidth]{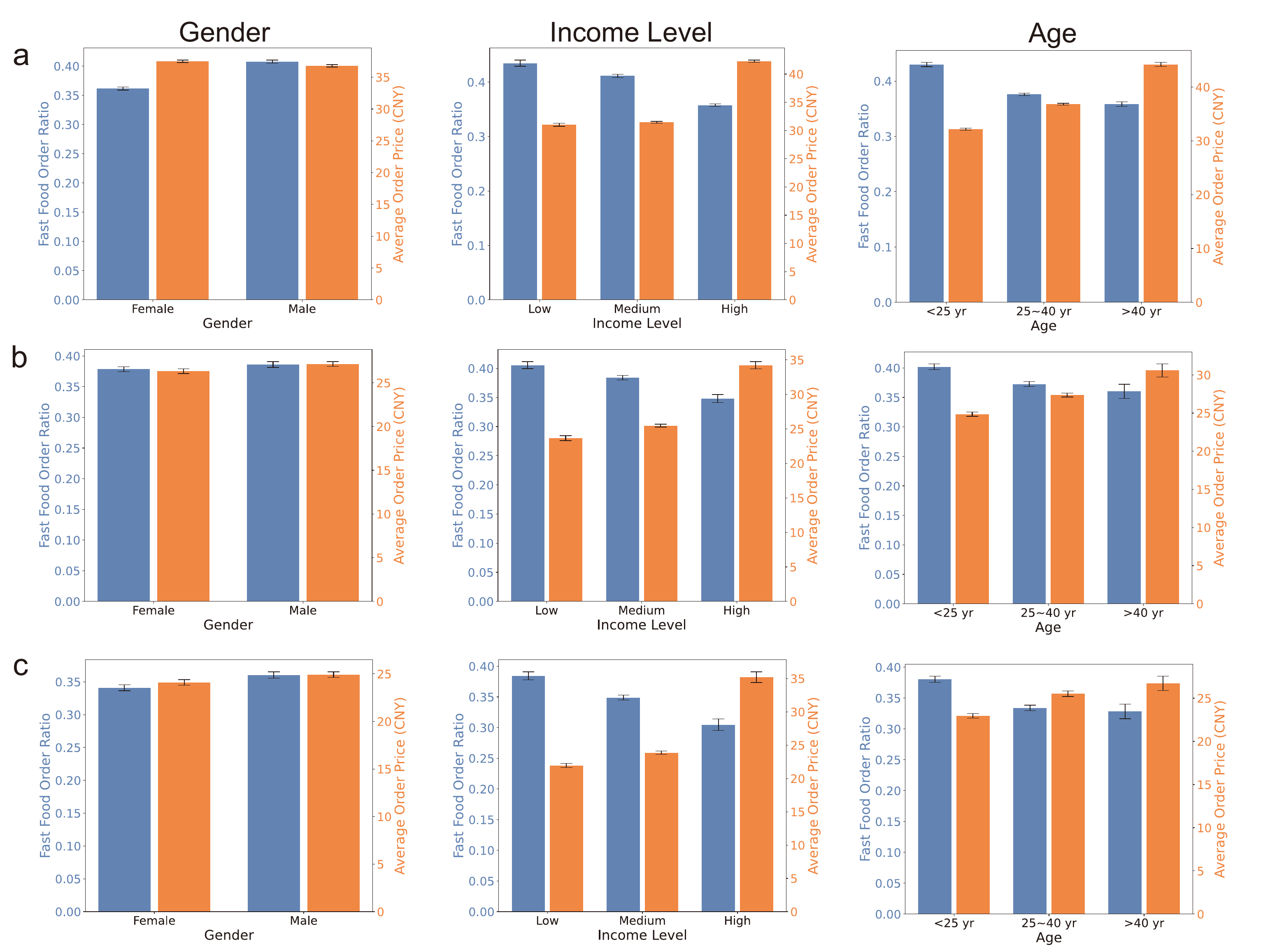}
\caption{The average fast food order ratio and order price across different gender, income level, and age groups in Beijing (a), Chengdu (b), and Xiamen (c) in 2024. Error bars represent 95\% confidence intervals.}
\label{fig:ineq_24}
\end{figure}

\begin{figure}[htbp]
\centering
\includegraphics[width=0.81\columnwidth]{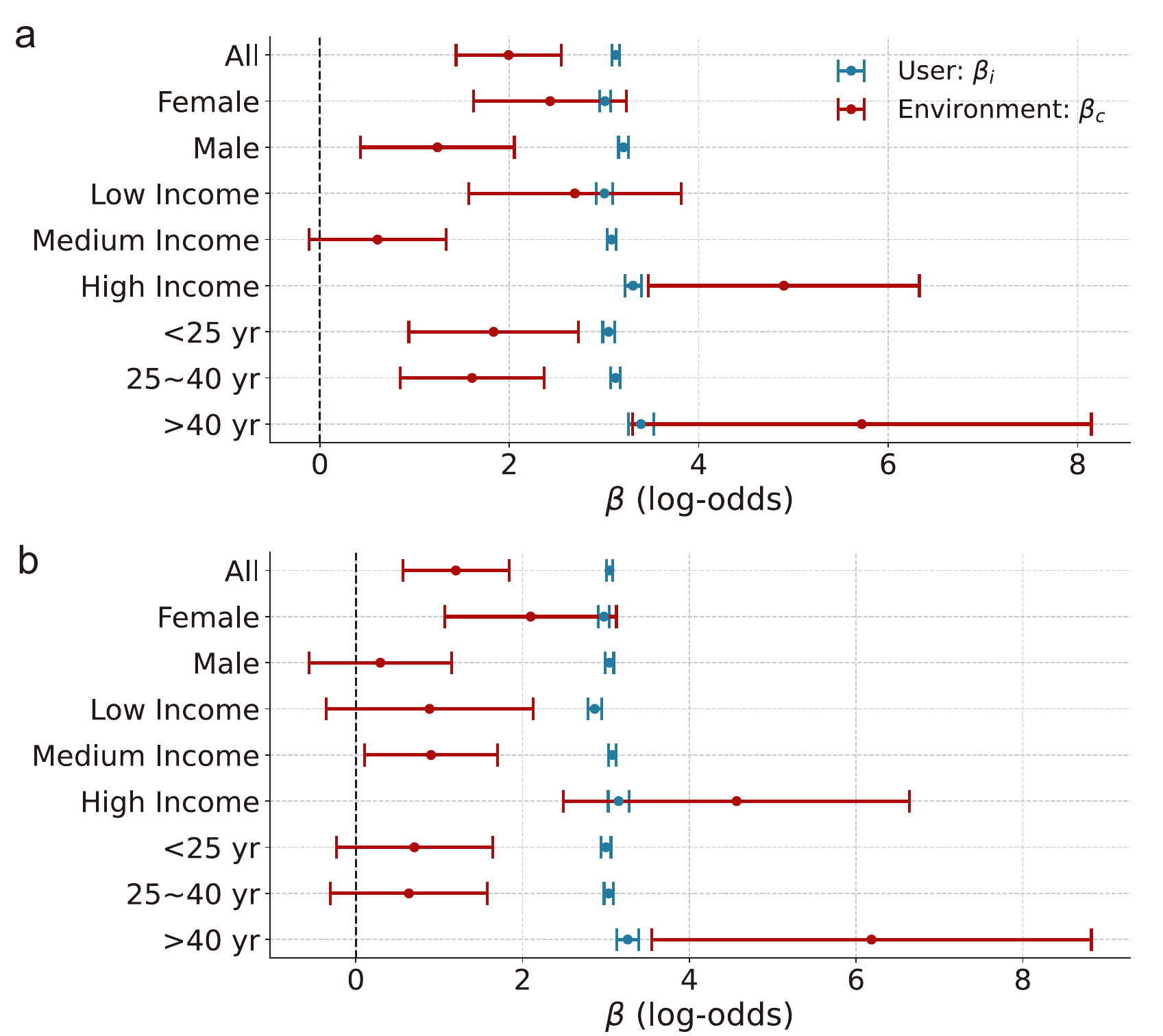}
\caption{Effect of user's historical preference ($\beta_i$) and online food environment ($\beta_c$) on the probability of placing a fast food delivery order over a non-fast food delivery order for different gender, income, and age groups in Chengdu (a) and Xiamen (b). Error bars represent 95\% confidence intervals for $\beta$'s.}
\label{fig:lr_appendix}
\end{figure}

\end{document}